# Leveraging Low-Fidelity Data to Improve Machine Learning of Sparse High-Fidelity Thermal Conductivity Data via Transfer Learning

Zeyu Liu[1,2], Meng Jiang[3], Tengfei Luo[1,4*]

[1]Department of Aerospace and Mechanical Engineering, University of Notre Dame, Notre Dame, Indiana, 46556, United States

[2]Department of Applied Physics, School of Physics and Electronics, Hunan University, Changsha 410082, China

[3]Department of Computer Science and Engineering, University of Notre Dame, Notre Dame, Indiana, 46556, United States

[4]Department of Chemical and Biomolecular Engineering, University of Notre Dame, Notre Dame, Indiana, 46556, United States

* Corresponding emails: tluo@nd.edu

**Abstract:** Lattice thermal conductivity (TC) of semiconductors is crucial for various applications, ranging from microelectronics to thermoelectrics. Data-driven approach can potentially establish the critical composition-property relationship needed for fast screening of candidates with desirable TC, but the small number of available data remains the main challenge. TC can be efficiently calculated using empirical models, but they have inferior accuracy compared to the more resource-demanding first-principles calculations. Here, we demonstrate the use of transfer learning (TL) to improve the machine learning models trained on small but high-fidelity TC data from experiments and first-principles calculations, by leveraging a large but low-fidelity data generated from empirical TC models, where the trainings on high- and low-fidelity TC data are treated as different but related tasks. TL improves the model accuracy by as much as 23% in $R^2$ and reduces the average factor difference by as much as 30%. Using the TL model, a large semiconductor database is screened, and several candidates with room temperature TC > 350 W/mK are identified and further verified using first-principles simulations. This study demonstrates that TL can leverage big low-fidelity data as a proxy task to improve models for the target task with high-fidelity but small data. Such a capability of TL may have important implications to materials informatics in general.

**Main**

Lattice thermal conductivity (TC) is a critical material property of semiconductors, impacting a wide range of applications, such as electronics thermal management,[1,2] thermoelectrics,[3,4] and thermal barrier coatings.[5] Low TC, for example, is of great interest for thermoelectric energy conversion, since the figure-of-merit is inversely proportional to TC. On the other hand, high TC is desirable for better electronics cooling to improve device reliability and performance. Identifying materials with desirable TC has mainly relied on physics-guided search, which is usually very time-consuming. Despite recent successes in achieving low[6–9] and high TC[10–14] in materials from physics-guided studies, the ability to fast-screen a large number of candidates for desired TC is highly valuable especially if design constraints are imposed on other properties (*e.g.*, low TC but high temperature stability for thermal barrier coating, and high TC with given lattice constants for electronics substrates). For developing new materials, the state-of-the-art method to predict TC is first-principles lattice dynamics combined with solving the phonon Boltzmann transport equation (BTE), which can be of high accuracy.[14–18] These calculations, however, can take a significant amount of time and computational resources as various parameters (e.g., pseudopotential selection, force constants cutoff, reciprocal space convergence, high-order phonon scattering, phonon renormalization) need to be carefully tuned to obtain reliable results. As a result, despite the high-fidelity of the first-principles calculations, the number of bulk semiconductor TC calculated using such a method is still limited.

Empirical methods can be leveraged for fast screening of the TC of a large number of semiconductors.[19] One example is using the Slack model with Debye temperature and Grüneisen parameters as input to predict TC. Recently, Toher et al. developed an Automatic Gibbs Library (AGL) to calculate the Debye temperature and Grüneisen parameters from first-principles, which are then used for high-throughput calculations of TC using the Slack model.[20] This effort has led to a relatively large dataset of semiconductor lattice TC with 1,567 materials, but the fidelity of such predicted TC is relatively low with apparent underestimation of the values.[20] However, since the AGL TC is calculated using a physical model, it should contain some meaningful composition-property relationships.

Data-driven machine learning (ML) methods provide another route to predict material TC.[21–23] Pioneer works includes a Bayesian optimization search based on 101 TC data from first-principles simulations for ultralow TC compounds by Seko et al.[24] Chen et al. later utilized experimentally available TC data for 100 semiconductors to train ML models and recursive feature elimination is applied to find the relevant descriptors.[25] However, the training data in these prior studies are relatively small, although with high fidelity. The small data can lead to biased coverage in the chemical space, and thus may restrict extrapolation capability of the trained ML models. For instance, there were only 8 materials out of 100 with TC over 100 W/mK in Chen's work,[25] potentially limiting the model's ability to screen high TC materials.

It would be ideal if one can leverage both the small but high-fidelity data and big but low-fidelity data to train ML models to increase the prediction accuracy and the range of applicability. Transfer learning (TL) is a ML scheme to overcome small data problems,[26] and it has been successfully applied in a few materials informatics studies. TL leverages knowledge based on a larger dataset of a proxy property and then transfers it to improve the ML model performance for a target property with a much smaller dataset.[23,27–31] In a previous study, Liu et al. used ~1400 electronic bandgap data from first-principles calculations to train a neural network, and its weights were used as initial parameters in training another neural network with 134 phonon bandgap data.[29] It was proven that the accuracy measured by mean absolute errors of prediction could be reduced by 65% for the target phonon property via such a TL strategy. It was noted that the first-principles electronic bandgap data were inaccurate due to the well-known density functional theory bandgap underestimation issue.[32] However, the first-principles calculations for phonon properties have been proven to have high-fidelity, and the trained TL model turned out to be highly accurate. Therefore, TL can be a potential technique to leverage big, low-fidelity data to help improve ML models trained with small but high-fidelity data.

In this work, we demonstrate the ability to use the TL strategy to improve the ML model accuracy trained with 170 high-fidelity TC data by leveraging 1,567 low-fidelity AGL TC as the proxy data. The 170 TC data combine experimentally measured values and those from first-principles lattice dynamics calculations from the literature. These two sets of data are different but related as they should both contain information of the structure-property relationship. Using the 1,567 AGL TC data, a deep neutral network (DNN) is first trained. TL is then applied to transfer the knowledge on the composition/structure-property relationship learned from the AGL data to the training using the 170 high-fidelity data (Fig. 1). TL is shown to improve the DNN model accuracy from $R^2 = 0.667$ to as high as 0.833 and reduces the average factor difference (AFD) from 2.18 to as low as 1.52. Using the TL model, all semiconductors in the *Materials Project* database[33] are screened, and several candidates with room temperature TC > 350 W/mK are identified, which are further verified using first-principles calculations. Our results demonstrate that TL can leverage big, low-fidelity data as a proxy to improve models for the target task with high-fidelity but small data. Such a capability of TL may have important implications to materials informatics in general.

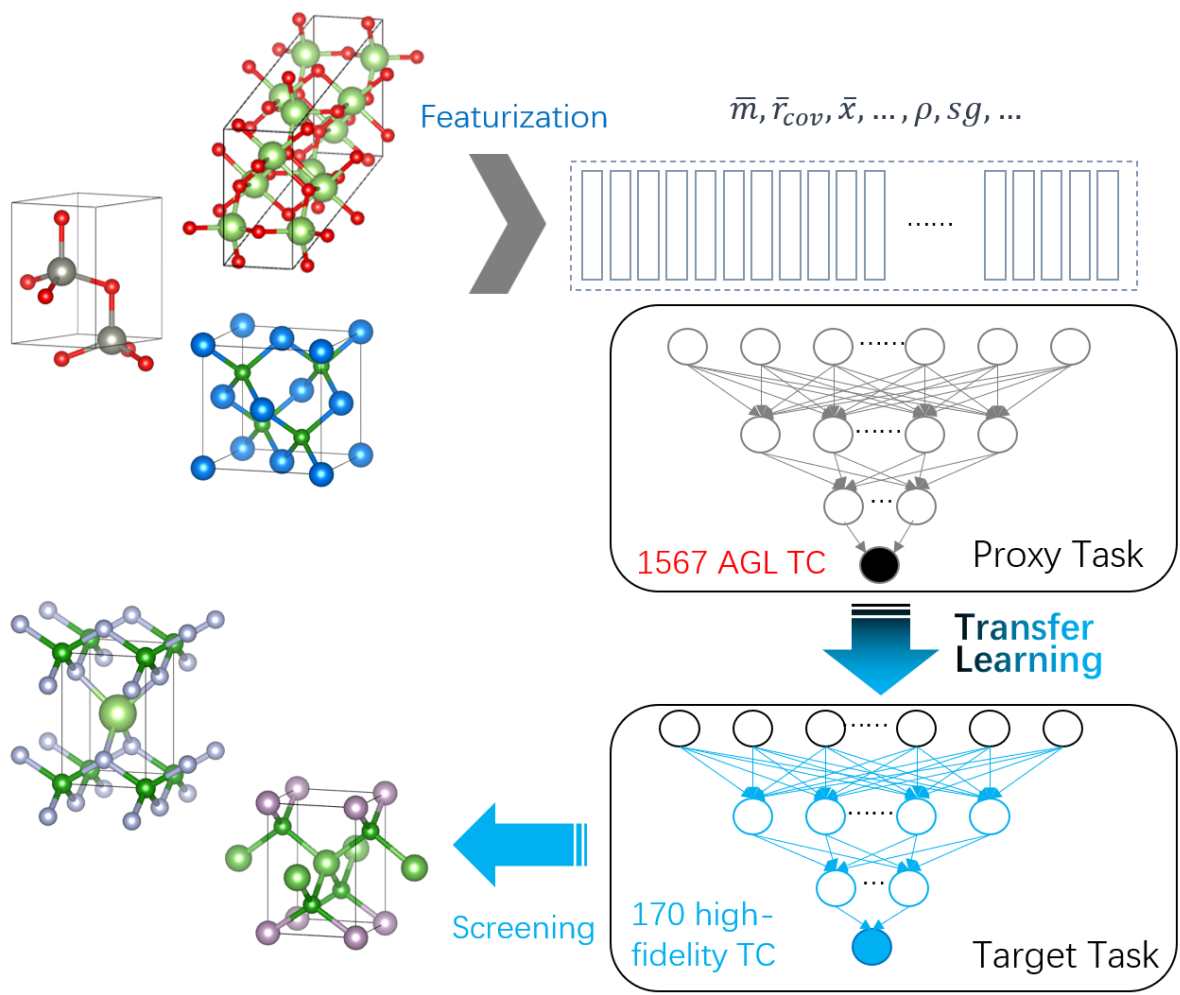

**Figure 1. Workflow of TL**. Schematics of the TL workflow using the big but low-fidelity AGL TC as the proxy task and transferring the learned knowledge to constructing a DNN model for the target task with high-fidelity but small data.

**Results**

For 101 semiconductors, TC values are available in both the AGL dataset and the high-fidelity dataset. We compare those TC values in a parity plot in Fig. 2a. The first observation is that there is a general positive correlation between the AGL data and the high-fidelity data, i.e., semiconductors with higher TC in reality also generally have higher TC in the AGL dataset. This suggests that the AGL TC dataset has encoded meaningful composition-property relation, which may be learnable by ML. However, AGL can underestimate the TC, especially for those with TC > 100 W/mK. On the other hand, the high-fidelity TC dataset spans over four orders of magnitude, even though with small data, indicating a sparse dataset (Fig. 2b). In comparison, the AGL dataset has a much narrower TC distribution centered around a low mean value (~3.5 W/mK) as shown in Fig. 2c. However, if we examine the representations of the semiconductors available in either dataset in a t-SNE plot (Fig. 2d), which visualizes the distribution of the high-dimensional data in a low-dimensional manifold,[34] it can be seen that the materials with high-fidelity TC cover a much smaller configuration space than those from the AGL dataset. Thus, building a ML model based on the small high-fidelity dataset will likely introduce large sampling bias. As a result, the ability to leverage the AGL dataset may help achieve ML models that are more generally applicable.

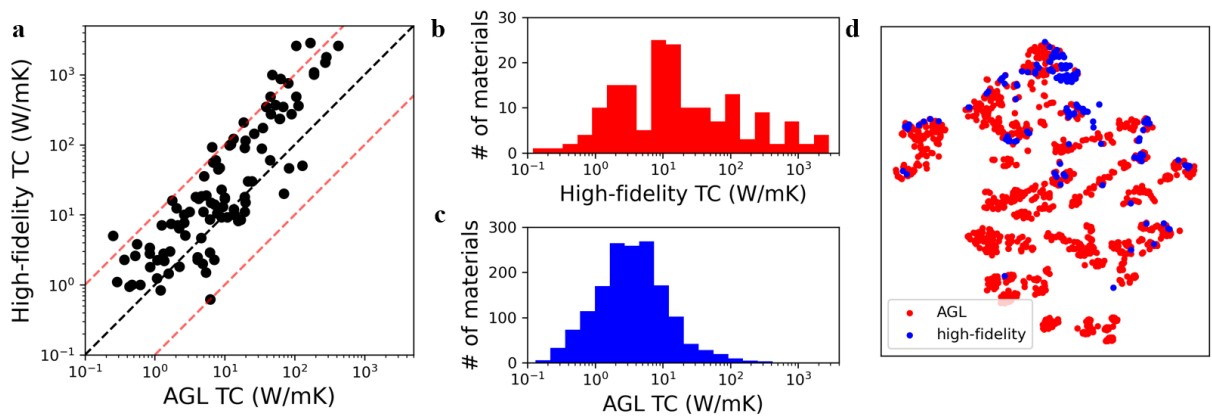

**Figure 2. Analysis of high- and low-fidelity datasets**. **a,** The comparison between the low-fidelity AGL TC dataset and the high-fidelity TC dataset from experiments and first-principles calculations. AGL can sometimes underestimate TC by more than one order of magnitude especially in the high TC region. **b,c,**

TC data distribution in the high-fidelity and AGL datasets. High-fidelity data is sparse, while the AGL is more narrowly distributed. **d,** A t-SNE visualization of the coverage of the configuration space by the descriptors of materials available in the high-fidelity dataset (red) and in the AGL dataset (blue). The AGL dataset covers a much larger configuration space than the high-fidelity dataset.

**ML model for proxy task:** Each semiconductor is represented using a 65-dimension vector which concatenates element-level compositional descriptors and crystalline structure descriptors (see Method). We first pre-train DNNs with the 1,567 AGL TC data with three hidden layers in a shot-gun approach,[28] the numbers of neurons in the layers of DNNs are randomized (see Method). A total of 91 shotgun models are separately trained with the AGL TC data using *PyTorch*[35]. The model regression performance metrics are evaluated in the five-fold cross-validation process – the predictions of each validation dataset are assembled and compared to the ground truth. For each model, 15 five-fold cross-validations each with a different random training-validation data split are performed. The total of 15 evaluated performance metrics are then averaged, and the standard deviation is used to account for the error bar. It should be noted that considering the wide span of orders of magnitude of TC, the logarithmic scale ($\log_{10} \kappa_L$) is usually used in place of the absolute TC value as the label for model performance assessment as did in other TC studies.[25,36–38] The average factor difference (AFD),[36] defined as AFD $= 10^a$, where $a = \frac{1}{n} \sum_n |\log_{10} \kappa_{predicted} - \log_{10} \kappa_{truth}|$ ($n$ is the index of the data point) is employed to quantify the average order of magnitude derivation of the model prediction from the ground truth. The DNN model prediction of the validation sets is shown against the AGL data in Fig. 3a, and a high prediction accuracy of $R^2$ is 0.848 obtained with the corresponding AFD calculated to be 1.32 (or equivalently the mean absolute error (MAE) of $\log_{10} \kappa_L$ is 0.121). It should also be noted that all $R^2$ values reported here are calculated after the TC values are converted from the log scale back to the linear scale, which we deem as a fairer way of characterizing the model accuracy. The model parameters, including weights and biases, of the DNNs trained on the AGL data are stored for further TL training.

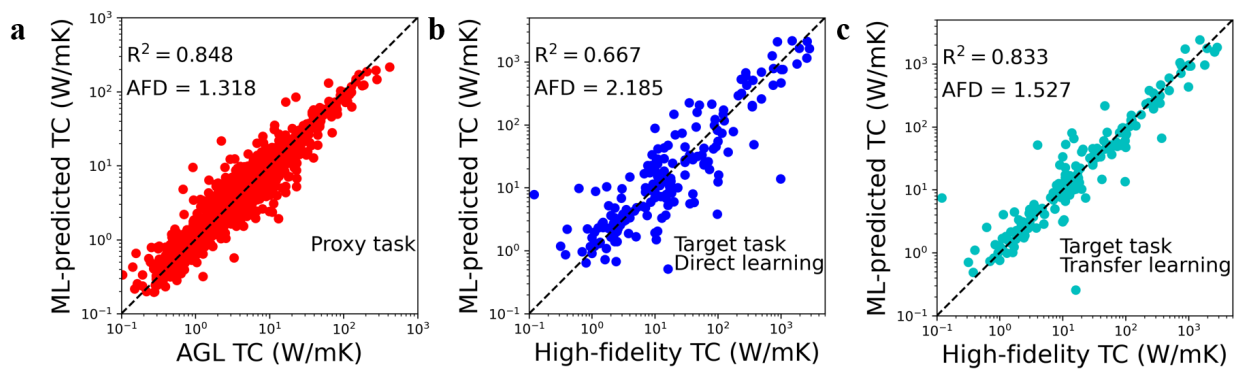

**Figure 3. ML model performances for different tasks. a,** Parity plot between ML-predicted TC and those in the AGL dataset. Note: all comparisons are for data in the validation sets of a five-fold validation task. **b,c,** Parity plots between the ML-predicted TC and the high-fidelity TC in the validation sets where the ML model is **(b)** directly trained by the high-fidelity data or **(c)** from TL.

**TL for target task:** For our target task, which is establishing ML models for the high-fidelity TC, we first directly train the DNN using the 170 high-fidelity data. DNNs with the same architectures as in the proxy task (i.e., the same 91 shotgun models created for the AGL data) are used, and the same five-fold cross-validation is used to evaluate model performance on the validation sets. Such a direct training led to a $R^2$ of 0.667 ± 0.684 and an AFD of 2.18 (Fig. 3b). We then apply TL in this target task, where we use the same DNN model architectures but with the weights and bias from the DNN trained in the proxy AGL TC task as the starting parameters – in contrast to the random parameters used in direct training. These DNN models are then re-trained using the small but high-fidelity TC data. Such a so-called warm-start TL strategy can greatly improve the $R^2$ and reduce the AFD. The $R^2$ can be improved by as much as 23% to 0.833, and the AFD can be reduced by as much as 30% to 1.52 (Fig. 3c). These results suggest that TL leveraging the AGL TC can indeed improve the ML model accuracy, and the knowledge embedded in the big AGL TC dataset, although of low-fidelity, can be useful for optimizing the ML model training on the small, high-fidelity TC dataset.

The performance of ML models can be influenced by the architecture and training/validation dataset distribution, especially for small data in our target task. We show the box plot of the $R^2$ and AFD metrics from all 91 shotgun cases in Fig. 4a, indicating that the increase in $R^2$ and decrease in AFD is statistically significant. Figure 4b shows the comparison of $R^2$ between direct learning and TL of all the 91 cases, which shows that TL can improve model accuracy in ~97% of the cases (points above the diagonal line). Figure 4c shows the comparison of AFD, which indicates TL can improve model accuracy in all the cases. In terms of $R^2$, the majority of the TL models are above 0.7, while the direct learning is hardly above 0.7. Similarly, direct training always leads to AFDs over 2.0, while TL models can improve AFD to 1.4 -1.6. Among the only three cases where TL does not outperform direct learning in $R^2$, two of them have similar performance metrics with only one case seeing obvious performance degradation from $R^2$=0.62 to 0.42. The observed effectiveness of the TL can be understood as that the training process based on the big AGL TC data can extract meaningful composition-property relationships embedded in the AGL data, and such knowledge is successfully transferred to the ML model trained by the high-fidelity data. When directly trained using the high-fidelity data, the low prediction accuracy on the validation set suggests that limited data restricts the generalizability of the trained model hence the low prediction accuracy for the validation sets. The bigger data and more diverse coverage of the configurational space (Fig. 2c) of the AGL dataset likely embeds more universal composition-property relationship which helps the TL model to be more generalizable.

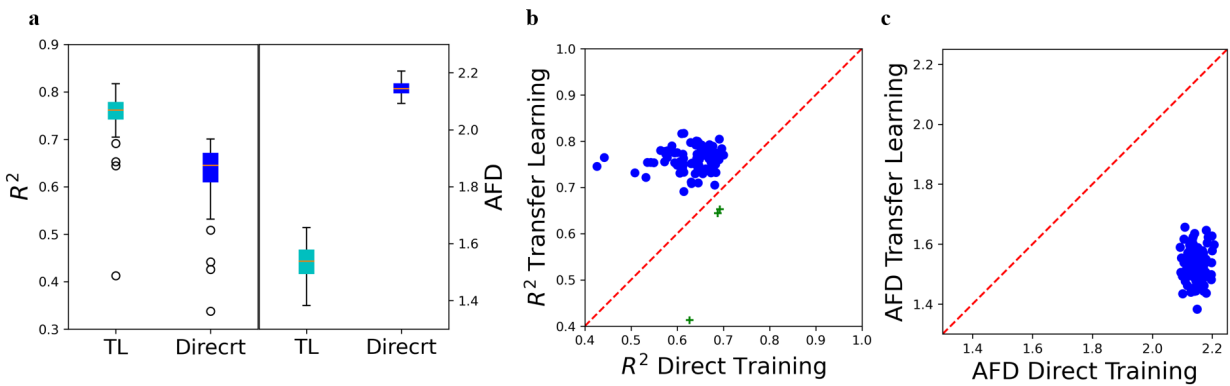

**Figure 4. Performance comparison between direct learning and TL models. a,** Boxplot comparison of the $R^2$ and AFD of the two types of learning with the statistics from different shotgun cases. **b,c,** Scattered plots of the **(b)** $R^2$ and **(c)** AFD from both types of learning for all shotgun cases.

One of the design tasks of interest to many applications is screening candidates to find semiconductors with relatively high TC (e.g., > 100 W/mK). A question arises is if one should include all available data in the target task, including low TC values, in the TL model, or one only needs high TC values to achieve a highly accurate model for the high TC region. Another shotgun TL experiment is performed using the same DNN architectures for the scenario where only the TC > 100 W/mK data in the high-fidelity dataset is used for TL. The same AGL dataset is used as the proxy task, and the trained TL models (denoted as $TL_{>100}$) performance is evaluated by using the models to predict the TC in the validation sets. As a comparison, we can also use the previous TL and direct learning models to predict the TC of the same group of semiconductors and evaluate their prediction accuracy. The distribution of the performance of different models is presented in Fig. 5. Although overlaps exist for the three types of models, TL using all available high-fidelity data can generally provide the best models, and TL based on only the high TC data are likely to yield less accurate models, which are still better than direct learning using all data. These observations suggest that low TC data is also helpful for high TC prediction, since it may embed useful information about composition-property relationships. As a result, it may be desirable to enlarge and diversify the data in the target task to maximize the TL model accuracy.

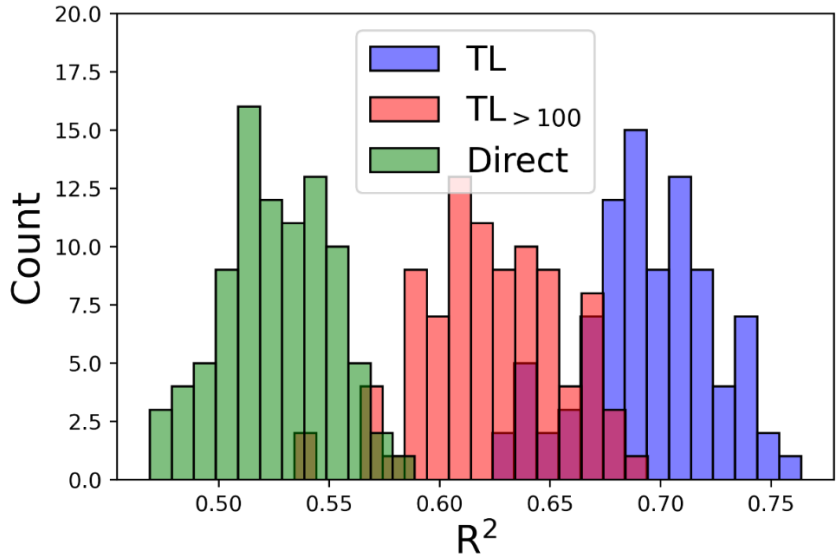

**Figure 5. TL strategy comparison for high TC prediction.** The histogram of the performance metrics of three different sets of models from direct learning, TL using all target task data and TL using target task data with TC > 100 W/mK.

**Screening high TC semiconductors:** We then use our obtained best TL model to identify possible high TC semiconductors. By screening all the semiconductors in the *Materials Project* database,[33] several materials are predicted to have high TC, and three examples are selected for validation using first-principles lattice dynamics TC calculations (see Methods). Two of the examples are $GaBN_2$ (mp-1007823)[39] and $AlBN_2$ (mp-1008557),[40] which share similar tetragonal structures with the space group number of 115. The crystal structures and the optimized lattice constants are shown in Fig. 6a. For $GaBN_2$, the TL-predicted TC is 540.28±56.83 W/mK, while the first-principles calculation predicts a TC of 387.90 W/mK for the x-direction and 169 W/mK for the z-direction (Fig. 6b). For $AlBN_2$, the TL model predicts its TC to be 533.78± 36.12 W/mK, while the first-principles-calculated TC is 510.30 W/mK in the x-direction and 164.06 W/mK in the z-direction. Another example tested is $B_2AsP$ (mp-1008528),[41] which has a tetragonal structure (Fig. 6a). The TL model predicts a TC of 611.77 ± 66.71 W/mK, while the first-principles calculation yields a TC of 371.57 and 307.84 W/mK along the x- and z-directions, respectively. As we can see, while the absolute values predicted by the TL models can differ from the first-principles-calculated ground truth, they are able to predict TC in the right orders of magnitude, and for $AlBN_2$, the TL prediction is very close to the ground truth in the x-direction. Characterized using the factor difference ($10^{|\log_{10} \kappa_{pred} - \log_{10} \kappa_{DFT}|}$), the errors for the three compounds are 1.39 ($GaBN_2$), 1.05 ($AlBN_2$), and 1.64 ($B_2AsP$) for their x-direction TC values, which are smaller or similar to the AFD of the TL models (Fig. 6c). Of course, the TL models are trained using TC data without crystal direction information, and thus understandably it cannot predict the TC anisotropy. In addition, the descriptors also do not consider the anisotropic crystal structural features. However, we believe these issues can be resolved with additional features and anisotropic data in the training set.

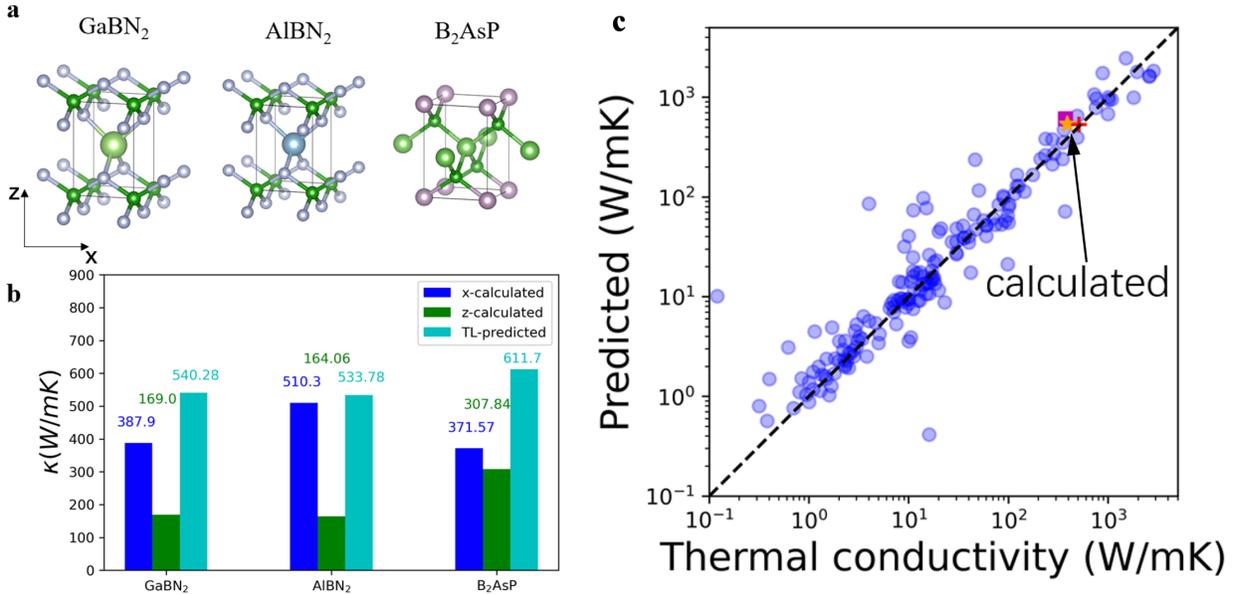

**Figure 6. Validation of TL-predicted high TC semiconductors**. **a,** Crystal structures of the three example compounds selected for first-principles validation. **b,** Bar plot of TL-predicted and first-principles-calculated TC values. **c,** The predicted three compounds shown in the parity plot of TL-prediction and high-fidelity TC data.

**Discussion and Conclusion**

We note that a semi-empirical Debye-Callaway model was able to achieve a AFD of ~1.5, within the error bar of our TL model, with the help of first-principles-calculated inputs for a different TC dataset.[36] The inputs include Debye temperature, speed of sound and Grüneisen parameters that are calculated from density functional theory[42]. It should be mentioned that there were fitting parameters in the model which thus also needed training. There were also *ad-hoc* additional terms (e.g., structural parameters) added to the Debye-Callaway model to improve the model accuracy. The benefit of data-driven models is that no prior knowledge of the physics is needed, and they can take only material information (e.g., elements, lattice structures) as inputs, without the needs of more complex descriptors from first-principles calculations. Of course, purely relying on data could lead to less accurate models, but our TL approach has been proven to solve such a problem.

The ML model prediction performance can vary, sometimes significantly, with the distribution of the dataset used for training, validation and testing. It also depends on how the performance metrics are evaluated. These factors make fairly comparing different models from different studies challenging. For example, Chen et al.[25] constructed a ML model using the more sophisticated recursive feature engineering and obtained a prediction $R^2 = 0.927$ for the test data. However, this metric was obtained from predicting a very small testing data size (5%) that is more uniformly distributed across the span of the TC values. For our best performing model on some certain training/validation data splits could also achieve a $R^2 = 0.97$. Zhu et al.[37] employed a Crystal Graph Convolutional Neural Networks (CGCNN) model[44] to predict TC and achieved $R^2 = 0.85$ on a reserved test data set, which is comparable to the $R^2$ in this work. However, we believe for small data, which is the case for TC, performing statistical study by randomizing the training/validation splits and DNN models in a shotgun approach should offer a more comprehensive assessment of the TL performance. In addition, our comparison between TL and direct training by testing them on exactly the same shotgun models and training/validation splits should have revealed the true benefit of TL. This benefit should also be applicable to more sophisticated ML models like the CGCNN.[44]

The most important finding from this study is that we can use low-fidelity but big data as a proxy to perform TL in order to improve the ML model accuracy trained using small but high-fidelity data. The low-fidelity data, although containing larger errors, still has valuable knowledge of composition-property relationships embedded in it, which is transferable to inform better training with much smaller high-fidelity data. Due to the more diverse configurational space coverage by the proxy data, transferring the composition-property relationship is helpful in improving the TL model when trained with the small high-fidelity data which covers much sparser configurational space. The implication of this study can be significant as it demonstrates the possibility of using data from simplified numerical or empirical models that can be generated in large volume as the proxy to inform the data-driven model trained on limited high-fidelity data that may be time- and resource-demanding to collect. Such a strategy should be able to benefit a wide range of applications for materials design.

**Methods**

**Material representation:** In order to represent the semiconductors in a fixed length vector for ML tasks, element-level compositional descriptors implemented by the *Matminer* project and precalculated by *MagpieData* dataset,[45] where a total of 61 elemental property descriptors, such as atomic weight, atomic radius and etc., are included. Besides elemental descriptors, several crystalline structure descriptors, including space group number, density, number of sites and volume per unit cell, are also included, making the total dimension of the representation 65. As in standard ML practices, all features are scaled in the range of 0 to 1 during model training. The high-fidelity thermal conductivity data is collected manually by searching both experimental and computational research for semiconductors at room temperature. And the low fidelity thermal conductivity data for semiconductor are collected from the Automatic Gibbs Library by restricting the bandgap to be exist.[20]

**DNN for AGL TC:** The DNN models are trained using *PyTorch*.[35] For the AGL TC (proxy property), there are three hidden layers in our DNN models, and the number of neurons in each hidden layer is randomly selected. The first two layers in a DNN contain a random number of neurons (1st layer: 60-90 neurons; 2nd layer: 20-50 neurons) and the 3rd layer contains a fixed 10 neurons. A total of 91 such different shotgun models are pre-trained, and the model performance metrics are calculated using the ground truth and the properties averaged by 15 independent five-fold cross-validation predictions. To evaluate the model prediction accuracy statistically, the predictions of each validation dataset in the five-fold cross-validation process are assembled and compared to the ground truth. The $R^2$ and AFD as the metrics of the 91 models are averaged and their standard deviations are reported as the error bar.

**Transfer learning (TL):** In the TL, we build DNN models with exactly the same architecture as the pre-trained models for the proxy property, and the parameters from these pre-trained models, except those for the output layer, are used as the initial parameters. These transferred models are then re-trained using the small data in the target task which includes high fidelity TC.

**First-principles calculations of TC:** For the first-principles calculations, two scattering mechanisms are included: three-phonon scattering and isotope scattering. A iterative process is employed to solve the phonon BTE to obtain the converged phonon distribution and TC at 300 K using *ShengBTE*.[46] For the three-phonon scattering calculation, a $6 \times 6 \times 6$ **q**-grid is adopted in the density functional perturbation theory[47] scheme to calculate the harmonic force constants using *Quantum Espresso*.[48] The third order force constants are calculated using a $4 \times 4 \times 4$ supercell with the 8th nearest neighbor cutoff via a finite difference method. The isotope scattering is calculated based on the natural isotopic distribution of the atoms using the Tamura equation.[49] These parameters are shown to yield converged TC values. It should be noted that for some materials at high temperatures, higher order phonon scattering like the four-phonon scattering can also play a role.[18,50,51] However, our first-principles validations are meant to provide an order of magnitude estimation of the TC for comparison with the TL model predictions. In addition, at the temperature of 300 K, we expect the effect of the four-phonon scattering to be relatively small.

**Data availability**

All data used in this study is available in the GitHub repository at https://github.com/liuzyzju/TL_thermal.

**Code availability**

All code from this study is available in the GitHub repository at https://github.com/liuzyzju/TL_thermal.

**Acknowledgements**

The computation is supported, in part, by the University of Notre Dame, Center for Researching Computing and NSF through XSEDE resources provided by TACC Stampede2 under grant number TG-CTS100078. Z.L. would like to thank the Fundamental Research Funds for the Central Universities (Grant No. 531118010723).

**Reference:**

1. Cui, Y., Li, M. & Hu, Y. Emerging interface materials for electronics thermal management: experiments, modeling, and new opportunities. *J Mater Chem C* **8**, 10568–10586 (2020).

2. Zheng, Q., Hao, M., Miao, R., Schaadt, J. & Dames, C. Advances in thermal conductivity for energy applications: a review. *Prog. Energy* **3**, 012002 (2021).

3. Tian, Z., Lee, S. & Chen, G. Comprehensive review of heat transfer in thermoelectric materials and devices. *Annu. Rev. Heat Transf.* **17**, (2014).

4. Toberer, E. S., Baranowski, L. L. & Dames, C. Advances in Thermal Conductivity. *Annu. Rev. Mater. Res.* **42**, 179–209 (2012).

5. Darolia, R. Thermal barrier coatings technology: critical review, progress update, remaining challenges and prospects. *Int. Mater. Rev.* **58**, 315–348 (2013).

6. Chiritescu, C. *et al.* Ultralow Thermal Conductivity in Disordered, Layered $WSe_2$ Crystals. *Science* **315**, 351–353 (2007).

7. Kim, S. E. *et al.* Extremely anisotropic van der Waals thermal conductors. *Nature* **597**, 660–665 (2021).

8. Mukhopadhyay, S. *et al.* Two-channel model for ultralow thermal conductivity of crystalline $Tl_3VSe_4$. *Science* **360**, 1455–1458 (2018).

9. Wang, S.-F., Zhang, Z.-G., Wang, B.-T., Zhang, J.-R. & Wang, F.-W. Intrinsic Ultralow Lattice Thermal Conductivity in the Full-Heusler Compound $Ba_2AgSb$. *Phys. Rev. Appl.* **17**, 034023 (2022).

10. Kang, J. S., Li, M., Wu, H., Nguyen, H. & Hu, Y. Experimental observation of high thermal conductivity in boron arsenide. *Science* **361**, 575–578 (2018).

11. Tian, F. *et al.* Unusual high thermal conductivity in boron arsenide bulk crystals. *Science* **361**, 582–585 (2018).

12. Li, S. *et al.* High thermal conductivity in cubic boron arsenide crystals. *Science* **361**, 579–581 (2018).

13. Chen, K. *et al.* Ultrahigh thermal conductivity in isotope-enriched cubic boron nitride. *Science* **367**, 555–559 (2020).

14. Lindsay, L., Broido, D. A. & Reinecke, T. L. First-Principles Determination of Ultrahigh Thermal Conductivity of Boron Arsenide: A Competitor for Diamond? *Phys Rev Lett* **111**, 25901 (2013).

15. McGaughey, A. J. H., Jain, A., Kim, H.-Y. & Fu, B. (傅博). Phonon properties and thermal conductivity from first principles, lattice dynamics, and the Boltzmann transport equation. *J. Appl. Phys.* **125**, 11101 (2019).

16. Lindsay, L. First Principles Peierls-Boltzmann Phonon Thermal Transport: A Topical Review. *Nanoscale Microscale Thermophys. Eng.* **20**, 67–84 (2016).

17. Esfarjani, K., Chen, G. & Stokes, H. T. Heat transport in silicon from first-principles calculations. *Phys. Rev. B* **84**, 085204 (2011).

18. Feng, T., Lindsay, L. & Ruan, X. Four-phonon scattering significantly reduces intrinsic thermal conductivity of solids. *Phys Rev B* **96**, 161201 (2017).

19. Plata, J. J. *et al.* An efficient and accurate framework for calculating lattice thermal conductivity of solids: AFLOW—AAPL Automatic Anharmonic Phonon Library. *Npj Comput. Mater.* **3**, 45 (2017).

20. Toher, C. *et al.* High-throughput computational screening of thermal conductivity, Debye temperature, and Grüneisen parameter using a quasiharmonic Debye model. *Phys. Rev. B - Condens. Matter Mater. Phys.* **90**, 1–14 (2014).

21. Sutton, C. *et al.* Crowd-sourcing materials-science challenges with the NOMAD 2018 Kaggle competition. *Npj Comput. Mater.* **5**, 1–11 (2019).

22. Pilania, G., Gubernatis, J. E. & Lookman, T. Multi-fidelity machine learning models for accurate bandgap predictions of solids. *Comput. Mater. Sci.* **129**, 156–163 (2017).

23. Ju, S. *et al.* Exploring diamondlike lattice thermal conductivity crystals via feature-based transfer learning. *Phys. Rev. Mater.* **5**, 053801 (2021).

24. Seko, A. *et al.* Prediction of Low-Thermal-Conductivity Compounds with First-Principles Anharmonic Lattice-Dynamics Calculations and Bayesian Optimization. *Phys. Rev. Lett.* **115**, 205901 (2015).

25. Chen, L., Tran, H., Batra, R., Kim, C. & Ramprasad, R. Machine learning models for the lattice thermal conductivity prediction of inorganic materials. *Comput. Mater. Sci.* **170**, 109155 (2019).

26. Pan, S. J. & Yang, Q. A Survey on Transfer Learning. *IEEE Trans. Knowl. Data Eng.* **22**, 1345–1359 (2010).

27. Kong, S., Guevarra, D., Gomes, C. P. & Gregoire, J. M. Materials representation and transfer learning for multi-property prediction. *Appl. Phys. Rev.* **8**, 021409 (2021).

28. Yamada, H. *et al.* Predicting Materials Properties with Little Data Using Shotgun Transfer Learning. *ACS Cent. Sci.* **5**, 1717–1730 (2019).

29. Liu, Z., Jiang Meng, & Luo Tengfei. Leverage electron properties to predict phonon properties via transfer learning for semiconductors. *Sci. Adv.* **6**, eabd1356.

30. Ma, R., Colón, Y. J. & Luo, T. Transfer Learning Study of Gas Adsorption in Metal–Organic Frameworks. *ACS Appl. Mater. Interfaces* **12**, 34041–34048 (2020).

31. Jha, D. *et al.* Enhancing materials property prediction by leveraging computational and experimental data using deep transfer learning. *Nat. Commun.* **10**, 5316 (2019).

32. Perdew, J. P. Density functional theory and the band gap problem. *Int. J. Quantum Chem.* **28**, 497–523 (1985).

33. Jain, A. *et al.* Commentary: The Materials Project: A materials genome approach to accelerating materials innovation. *APL Mater.* **1**, 11002 (2013).

34. Pedregosa, F. *et al.* Scikit-learn: Machine Learning in Python. *J. Mach. Learn. Res.* **12**, 2825–2830 (2011).

35. Paszke, A. *et al.* PyTorch: An Imperative Style, High-Performance Deep Learning Library. in *Advances in Neural Information Processing Systems 32* 8024–8035 (Curran Associates, Inc., 2019).

36. Miller, S. A. *et al.* Capturing Anharmonicity in a Lattice Thermal Conductivity Model for High-Throughput Predictions. *Chem. Mater.* **29**, 2494–2501 (2017).

37. Zhu, T. *et al.* Charting lattice thermal conductivity for inorganic crystals and discovering rare earth chalcogenides for thermoelectrics. *Energy Environ. Sci.* **14**, 3559–3566 (2021).

38. Jaafreh, R., Kang, Y. S. & Hamad, K. Lattice Thermal Conductivity: An Accelerated Discovery Guided by Machine Learning. *ACS Appl. Mater. Interfaces* **13**, 57204–57213 (2021).

39. Persson, K. Materials Data on GaBN2 (SG:115) by Materials Project. (2016) doi:10.17188/1324736.

40. Persson, K. Materials Data on AlBN2 (SG:115) by Materials Project. (2016) doi:10.17188/1325075.

41. Merabet, M. *et al.* Electronic structure of (B P)n/(B As)n (0 0 1) superlattices. *Phys. B Condens. Matter* **406**, 3247–3255 (2011).

42. Yan, J. *et al.* Material descriptors for predicting thermoelectric performance. *Energy Env. Sci* **8**, 983–994 (2015).

43. Zhang, Y. & Ling, C. A strategy to apply machine learning to small datasets in materials science. *Npj Comput. Mater.* **4**, 25 (2018).

44. Xie, T. & Grossman, J. C. Crystal Graph Convolutional Neural Networks for an Accurate and Interpretable Prediction of Material Properties. *Phys. Rev. Lett.* **120**, 145301 (2018).

45. Ward, L. *et al.* Matminer: An open source toolkit for materials data mining. *Comput. Mater. Sci.* **152**, 60–69 (2018).

46. Li, W., Carrete, J., Katcho, N. a. & Mingo, N. ShengBTE: A solver of the Boltzmann transport equation for phonons. *Comput. Phys. Commun.* **185**, 1747–1758 (2014).

47. Baroni, S., De Gironcoli, S., Dal Corso, A. & Giannozzi, P. Phonons and related crystal properties from density-functional perturbation theory. *Rev. Mod. Phys.* **73**, 515–562 (2001).

48. Giannozzi, P. *et al.* QUANTUM ESPRESSO: A modular and open-source software project for quantum simulations of materials. *J. Phys. Condens. Matter* **21**, (2009).

49. Tamura, S. Isotope scattering of large-wave-vector phonons in GaAs and InSb: Deformation-dipole and overlap-shell models. *Phys Rev B* **30**, 849–854 (1984).

50. Feng, T. & Ruan, X. Quantum mechanical prediction of four-phonon scattering rates and reduced thermal conductivity of solids. *Phys Rev B* **93**, 045202 (2016).

51. Han, Z., Yang, X., Li, W., Feng, T. & Ruan, X. FourPhonon: An extension module to ShengBTE for computing four-phonon scattering rates and thermal conductivity. *Comput. Phys. Commun.* **270**, 108179 (2022).

**Contributions**

All authors designed the research study and developed the method. Z.L. wrote the code and performed the analysis. All authors wrote and approved the manuscript.

**Ethics declarations**

Competing interests

The authors declare no competing interests.